\documentclass[aps,prl,preprint,superscriptaddress,showpacs]{revtex4-1}

\pdfoutput=1
\usepackage{gensymb}
\usepackage{amsmath}
\usepackage{graphicx}
\usepackage{epsfig}
\usepackage{multirow}

\begin{document}

\title{The Refraction of Surface Plasmon Polaritons}

\author{Jonathan J. Foley, IV}
\affiliation{Center for Nanoscale Materials, Argonne National Laboratory, Argonne, IL 60439}

\author{Jeffrey M. McMahon}
\affiliation{Department of Physics, University of Illinois at Urbana-Champaign, Urbana, IL 61801}

\author{George C. Schatz}
\affiliation{Department of Chemistry, Northwestern University, Evanston, IL 60208}

\author{Stephen K. Gray}
\affiliation{Center for Nanoscale Materials, Argonne National Laboratory, Argonne, IL 60439}
\email[]{gray@anl.gov}

\date{\today}

\begin{abstract}

We show how a complex Snell's law can be used to describe the refraction of 
surface plasmon polaritons (SPPs) at an interface between
two metals, validating its predictions with 3-D electrodynamics simulations.
Refraction gives rise to peculiar
SPP features including inhomogeneities in the waveform
and dispersion relations that depend on the incident
wave and associated material.  These features make it possible to generate SPPs
propagating away from the interface
with significant confinement normal to the propagation direction. 
We also show that it is possible to encode
optical properties of the incident material into the refracted SPP.
These consequences of metal-metal SPP refraction provide new
avenues for the design of plasmonics-based devices.

\end{abstract}

\maketitle


Surface plasmon polaritons (SPPs), surface waves created by coupling light into charge-density oscillations at a 
metal--dielectric interface, continue to be of current 
interest \cite{barnes-rev, zayats-rev, 
gramotnev-rev, S_OptExp_2011, BD_NatPhot_2012,Novotny, SKG_JPCC_2013}.
Systems that permit the excitation of SPPs can exhibit interesting and  
unexpected optical properties, including extraordinary optical transmission \cite{EOT_Ebbesen_1998} and  
super-lensing \cite{perfect-lens_Pendry-PRL-2000,FLS_Sci_2005, S_PRL_2007}.  Such properties  are also relevant to a wide range of applications 
including imaging and sensing \cite{McMahon_RASPP_Orig, McMahon_RASPP_params, stewart-chem-rev,CR_Homola, SPL_SciRep} and 
optoelectronics \cite{plasmonics-for-electronics_Ozbay-Science-2006, Xu_JPD, PSB_ACSNano_2012, SD_PRB_2012}. 
Therefore, learning how to control and manipulate SPPs as 
they propagate along a metal surface is a major goal of nanophotonics research. 
A particular goal for optoelectronics applications involves
introducing lateral confinement of the SPPs without sacrificing 
the propagation length \cite{APL_2007,NL_9, PSB_ACSNano_2012, NL_13}.
For sensing applications, sensitivity figures of merit depend not on the features of the SPP wave itself,
but on the SPP dispersion relations, where a desirable feature is a strong dependence
of the SPP dispersion on the dielectric environment \cite{CR_Homola,SPL_SciRep}. 

SPPs propagate on a 2-D metal surface, and are exponentially confined both above and below this surface, suggesting 
that one can attempt to describe and manipulate their motion 
using ideas from classical optics applied to the 2-D propagation plane. 
Successful examples include the focusing of SPPs
created by an array of holes or slits in
metal films \cite{yin-focus,bahns-sers} and the generation of Talbot effect
intensity patterns \cite{abajo-talbot,exp-talbot-1,exp-talbot-2}. SPPs have also been shown experimentally to
exhibit refraction behavior when they propagate across an interface
between two metal/dielectric interfaces with differing optical properties \cite{snell-1,snell-2}.
Negative refraction of SPP-dominated waveguide modes has also been achieved \cite{LDA_Science}. 
In this Letter, we analyze SPP refraction, presenting and discussing the implications of a
complex generalization of Snell's law (CSL).  The CSL predicts that 
refracted SPPs will be inhomogeneous and will obey dispersion relations
that depend not only on the medium supporting
the refracted SPP, but on the incident wave and medium.  The inhomogeneous character of the refracted wave can be exploited
to introduce significant confinement of the SPP without sacrificing
propagation length.  The dependence of the dispersion of the refracted
SPP on the details of the incident wave introduces the possibility of 
encoding or imprinting the plasmonic properties of one
material onto another, including anomalous dispersion phenomena (back-bending)
and `slow-light' \cite{PRL_Ritchie_1973}.  
Dispersion encoding imparts an unexpected environmental
sensitivity to the refracted SPP, 
which may be useful
for sensing applications.\cite{CR_Homola,Xu_JPD,SPL_SciRep}  

Refraction of 3-D plane waves from a dielectric medium into an absorbing medium
is a well-known problem \cite{Ciddor_1976,Parmigiani_1982,BornWolf} and 
there are also treatments of refraction that allow for the
medium of the incident wave to be absorbing \cite{Mahan_1956,Dupertuis_1994,chang_2005}.
Here we adapt the particularly transparent treatment of
Chang, Walker and Hopcraft\cite{chang_2005} to the case of 2-D SPP refraction at the
boundary between different metal surfaces
and discuss several consequences. 
The system of interest involves an SPP,
generated by conventional means, propagating on
top of a metal surface 1 that  has  dielectric material 1 above it; it is reflected and refracted
at the interface with a different metal surface 2 with  possibly different dielectric 2
above it (Fig. 1).  The refracted SPP  propagates on the surface of metal 2.
The propagation of the SPPs on each surface can be described with 2-D waveforms
moving in the $x-y$ plane of Fig. 1,
\begin{equation}
{\bf E}_j = {\bf E}_{0,j} \: {\rm exp}\left(i \: {\bf k}_j \cdot {\bf r} - i \omega t \right),
\end{equation} 
where ${\bf k}_j$ is a complex SPP wavevector associated with incident ($j$ = 1) or refracted
($j$ = 2) waves.  (For the present purposes reflection is not of relevance.)
A 2-D medium refractive index may be defined
based on the standard SPP dispersion relation:  
\begin{equation}
\eta_j + i \kappa_j = \left( \frac{ \epsilon_j \epsilon^D_j}{\epsilon_j + \epsilon^D_j} \right)^{1/2},
\end{equation}
where $ \epsilon_j = \epsilon_j(\omega)$ is the frequency-dependent permittivity
of metal $j$ and $ \epsilon^D_j$ is the permittivity of the dielectric
material above it.  
%
\begin{figure}
\includegraphics[width=3.5in]{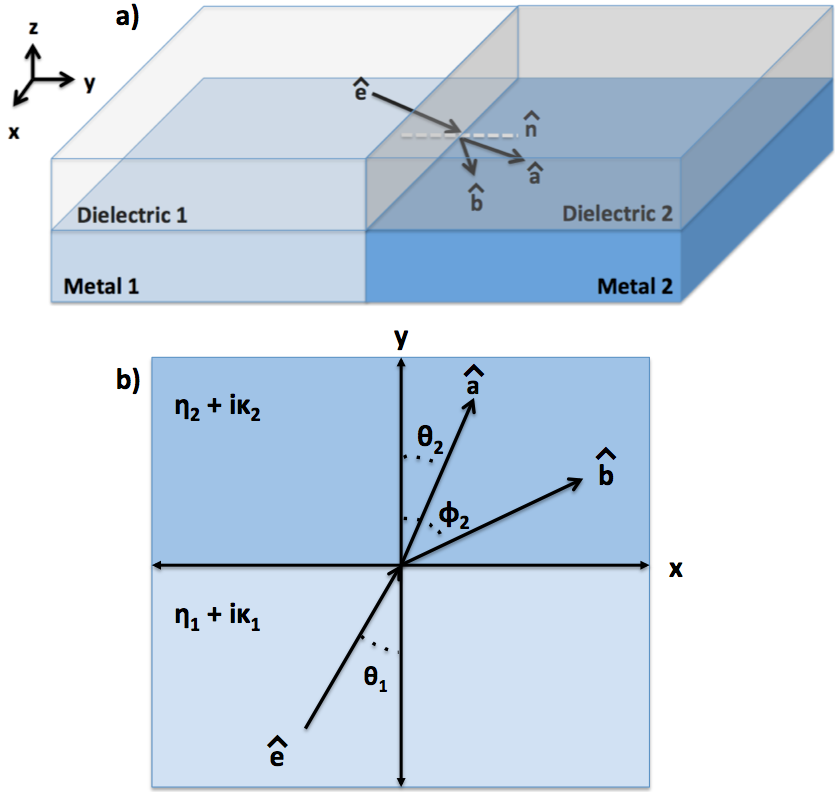}
  \caption{(a) Diagram of SPP refraction at a
metal-dielectric/metal-dielectric interface.  The incident
SPP propagates on the 2-D interface between metal 1 and dielectric 1,
with direction
$\hat{{\bf e}}$ being associated with both the real and
imaginary parts of its wavevector.  The wavevector of the refracted SPP propagating in the
interface region of metal 2 and dielectric 2
has direction $\hat{{\bf a}}$ associated
with its real part and direction $\hat{{\bf b}}$
associated with its imaginary part. (b) 2-D effective medium 
picture of the refraction in the $x-y$ plane of (a).}
  \label{fig:Snell_schematic}
\end{figure}
The incident SPP wavevector is 
\begin{equation}
{\bf k}_1 = \frac{\omega}{c}\left( \eta_{1} + i \kappa_{1} \right) \hat{{\bf e}},
\end{equation}
where $\hat{{\bf e}} = {\rm sin}(\theta_1) \hat{x} + {\rm cos}(\theta_1) \hat{y}$
is a unit vector indicating the direction of propagation. 
The lines of
constant phase and constant amplitude for ${\bf E}_1$ are parallel and
$\hat{{\bf e}}$ is the direction normal to both these types of lines. 
${\bf E}_1$ is therefore homogeneous. 
In contrast, the refracted SPP is allowed to be inhomogeneous;
its lines of constant phase and amplitude are not necessarily parallel.  
The wavevector for the refracted SPP is thus taken to be
\begin{equation}
{\bf k}_2 = \frac{\omega}{c}\left( N_{2} \: \hat{{\bf a}} + i K_{2} \: \hat{{\bf b}} \right)
\end{equation}
where $\hat{{\bf a}} = {\rm sin}(\theta_2) \hat{x} + {\rm cos}(\theta_2) \hat{y}$
is a unit vector normal to the lines of constant phase,
$\hat{{\bf b}} = {\rm sin}(\phi_2) \hat{x} + {\rm cos}(\phi_2) \hat{y}$
is a unit vector normal to the lines of constant amplitude,
and the effective indices $N_2$ and $K_2$ depend on the medium
refractive indices $\eta_2$, $\kappa_2$, $\eta_1$, $\kappa_1$ and
incident angle, $\theta_1$.  
The boundary conditions and wave equation 
determine 
$N_2$ and $K_2$ in terms of the known quantities.
The SPP phases ${\bf k}_j \cdot {\bf r}$ must be continuous
at the $y$ = 0 metal-metal interface, 
implying
\begin{align}
\eta_1 \: {\rm sin}(\theta_1) &= N_2 \: {\rm sin}(\theta_2) \\
\kappa_1 \: {\rm sin}(\theta_1)   &= K_2 \: {\rm sin}(\phi_2)~~.
\end{align}
We refer to Eqs. (5) and (6) as the complex Snell's law (CSL) because they 
determine the angles of refraction of the real and imaginary parts of the complex refracted SPP wavevector.
All the quantities involved in these equations are still real
and so the complication of complex angles is avoided.
To determine $N_2$ and $K_2$, 
inserting ${\bf E}_2$ of the form Eq. (1) into the usual second-order electromagnetic
wave equation gives
\begin{equation}
\left({\bf k}_2 \cdot {\bf k}_2\right) {\bf E}_2 = \frac{\omega^2}{c^2}\left(\eta_2 + i\kappa_2\right)^2 {\bf E}_2,
\end{equation}
which is satisfied if 
\begin{equation}
N_2^2 - K_2^2 = \eta_2^2 - \kappa_2^2 \\
\end{equation}
and
\begin{equation}
N_2  K_2 \: {\rm cos}(\theta_2 - \phi_2) = \eta_2 \kappa_2.
\end{equation}
Using Eqs. (5) and (6), Eq. (9) is equivalent to 
\begin{equation}\label{eq:Quartic}
 \eta_2 \kappa_2 - \alpha_1 \beta_1 =  
\sqrt{N_2^2 - \alpha_1^2}\sqrt{N_2^2 - \left( \eta_2^2 - \kappa_2^2\right) - \beta_1^2} ,
\end{equation}
where $\alpha_1 = \eta_1 {\rm sin}(\theta_1)$ and $\beta_1 = \kappa_1 {\rm sin}(\theta_1)$.
Squaring both sides of Eq.~\ref{eq:Quartic} gives a quartic equation for $N_2$, which has
the following root of interest
\begin{equation}\label{eq:Root}
N_2 = \frac{1}{\sqrt{2}} \sqrt{ a + \sqrt{b} },
\end{equation}
where
\begin{align*}
a = & \alpha_1^2 + \beta_1^2 + \eta_2^2 - \kappa_2^2, \\
b = &\left( \left(\kappa_2 - \beta_1\right)^2 + \left(\eta_2 - \alpha_1\right)^2\right)\left( \left(\kappa_2 + \beta_1\right)^2 + \left(\eta_2 + \alpha_1\right)^2\right)
\end{align*}
Once $N_2$ is known, $K_2$ may be determined readily using
Eq. (8); $\theta_2$ and $\phi_2$ are then found
from form the CSL, Eqs. (5) and (6).  

In the case of normal incidence ($\theta_1$ = 0) 
the CSL 
leads to
$N_2$ = $\eta_2$,
$K_2$ = $\kappa_2$, and $\theta_2$ = $\phi_2$, i.e. a refracted SPP 
in medium 2 that is an ordinary medium 2 SPP.
Otherwise, there are two new features:  (1) $\theta_2$ $\neq$ $\phi_2$,
i.e. the waveform is inhomogeneous with its lines of constant phase and
constant amplitude no longer being parallel and (2) $N_2 + i K_2$ $\neq$ $\eta_2 + i \kappa_2$,
i.e. the complex propagation constant (and dispersion relation) is not the same as that for
an ordinary SPP in medium 2.

The propagation length  ($L_P$) of an SPP is the distance, measured
along the propagation direction, that the SPP propagates
when the intensity decays to $|{\bf E}|^2_0/e$. For an ordinary (homogeneous) SPP
in medium 2, $L_P = 1/(2 k_0 \kappa_2)$, where
$k_0$ is the free-space wavevector of the exciting light, $k_0 = 2\pi/\lambda_0$. 
For a refracted
SPP in medium 2, this distance is measured along $\hat{{\bf a}}$ and
is given by $L_P = 1/(2 k_0 K_2 {\rm cos}(\theta_2 - \phi_2))$. 
Utilizing Eq. (9), the ratio of the refracted SPP propagation length to an
ordinary SPP propagation length is simply $N_2/\eta_2$ and so if $N_2 > \eta_2$
there will be propagation length enhancement. 
We define a {\it confinement length} ($L_C$) as the 
distance in the direction perpendicular to propagation over which the SPP intensity
decays to  $|{\bf E}|^2_0/e$.  The confinement length of an inhomogeneous SPP
is given by $L_C = 1/(2 k_0 K_2 {\rm sin}(\theta_2 - \phi_2))$, where
a more strongly confined SPP has a shorter confinement length.
In the event of propagation length enhancement, $K_2$ itself
will be larger than $\kappa_2$; hence propagation length enhancement is also
associated with strong confinement of the refracted SPP.

\begin{figure}
\includegraphics[width=5.5in]{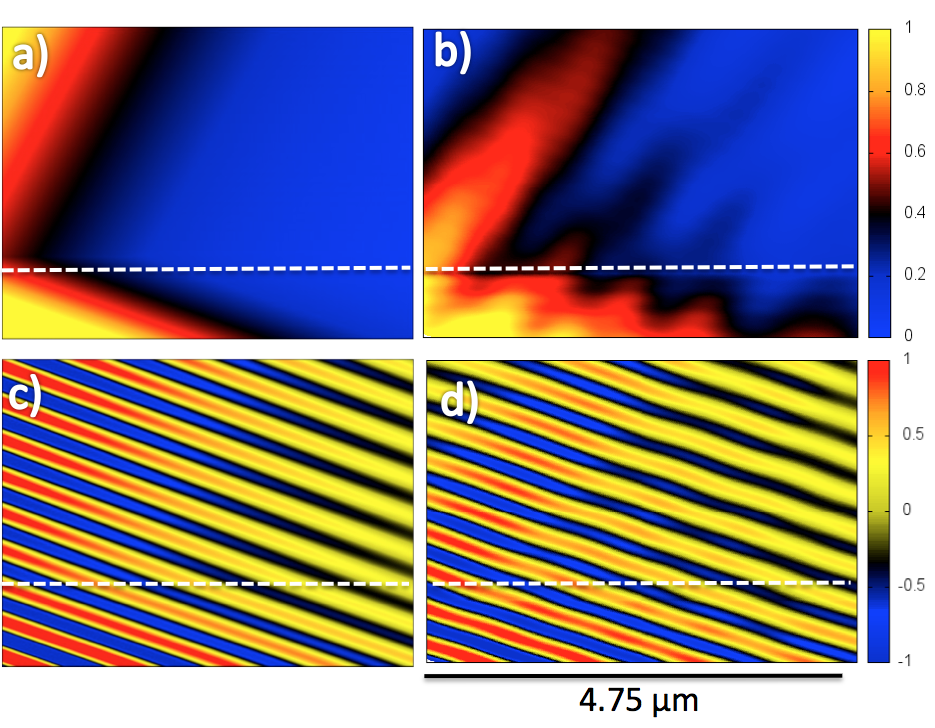}
\caption{  
Field profiles of incoming SPP (below
white dashed line) excited on an Au surface by
532 nm light with incident angle 25$^{\circ}$ refracting onto an Ag 
surface (above white dashed line).  Air is assumed to be above both surfaces.
(a) and (b): Analytical CSL and FDTD electric field intensities,
respectively.  (c) and (d): Analytical and FDTD instantaneous
electric field components, respectively. The FDTD results are 
$x-y$ cuts taken at a $z$ level 120 nm above metal surfaces and are associated
with the $z$ component of the electric field.}
\label{fig:AuAg_Montage_Ez}
\end{figure}

We first consider Au for metal 1 and Ag for metal 2 with air
as the dielectric above each metal, and an incident SPP on Ag
excited with $\lambda_0$ = 532 nm light.
We take $\epsilon_{1} = -4.762 + 2.378 \; i$
and
$\epsilon_{2} = -11.825 + 0.374 \; i$\cite{JC_PRB_1972}
which gives medium refractive indices 
$\eta_1 = 1.092$, $\kappa_1 = 0.057$,
$\eta_2 = 1.045$, and $\kappa_2 = 0.0015$.
While $\eta_1$ and $\eta_2$ are similar in magnitude,
$\kappa_1$ is significantly larger than $\kappa_2$
and so $\phi_2$ rises very rapidly with $\theta_1$ (see Fig. S3).
When an incident angle of 25$^{\circ}$  is considered, 
the CSL predicts $\theta_2$ to be 26$^{\circ}$ and $\phi_2$ to be 
112$^{\circ}$ (see Fig. S3).  ($\phi_2$ can be larger than
90$^{\circ}$ because the arcsine has two unique values, a principal value
between 0 and $\pi/2$ and a secondary value between $\pi/2$ and $\pi$; 
one or the other of these values is the physically correct one.
See the Supporting Material)
The analytical CSL electric field intensity ($| {\bf E}_z | ^2$) 
and instantaneous field map, ${\rm Re}({\bf E}_z)$,  
are shown in Fig. 2 (a) and (c), respectively.
To validate these predictions, we use rigorous 3-D 
finite-difference time-domain (FDTD) 
calculations \cite{Taflove_FDTD,MLG_FDTD} (see Supporting Material) to simulate
the refraction phenomena and plot $| {\bf E}_z | ^2$ and
${\rm Re}( {\bf E}_z )$ in Fig. 2 (b) and (d), respectively.
(Other FDTD field components give similar results.)
The FDTD results show 
a high degree of similarity in the SPP wavelength, 
propagation direction, and attenuation behavior with that predicted by CSL.
Values of the simulated electric field
intensity are sampled along the propagation direction (${\bf \hat{a}}$) and fit an exponential
to allow accurate inference of $L_P$ (see Fig. S1).
The FDTD fields do differ from the CSL ones in that there is a fast decay of
the FDTD field in the upper left region of Figs. 2 (b) and (d).
This is simply due to 
the finite size of the excitation source used in the simulations 
(see Supplementary Material). One can also see slight interference fringes
in the FDTD results due to interference of incident and reflected waves.

We find that the propagation length of the refracted
SPP as determined by rigorous 3-D electrodynamics calculation is 
approximately 28 $\mu$m, compared to 27 $\mu$m 
from analytical predictions using CSL.  The FDTD instantaneous
field allows us to infer a propagation direction of 27$^{\circ}$,
compared to the CSL prediction of 26$^{\circ}$.
The simulated propagation length thus
closely matches the propagation length predicted for a normal
SPP excited on a silver surface, but the refracted SPP has
significant lateral confinement compared to an ordinary SPP.
We follow a similar procedure to extract $L_C$, this time sampling
the electric field intensity along the direction perpendicular to ${\bf \hat{a}}$,
and nearly parallel to ${\bf \hat{b}}$.
We find $L_C$ of the refracted SPP is 1.1 $\mu$m as determined
by FDTD calculation, which agrees closely to the CSL prediction
of  1.7 $\mu$m (see Fig. S1).  
Note that an ordinary SPP in this medium would propagate with {\it no}
such confinement lateral to the propagation direction.
Analysis of the FDTD lines
of constant amplitude allows us to infer an attenuation direction of 115$^{\circ}$,
similar to the CSL prediction of 112$^{\circ}$.
In this first example, $\phi_2$ is relatively
large relative relative to $\theta_2$, i.e., the direction of the amplitude 
decay is nearly normal to the propagation direction, a dramatic change
relative to the incident wave that had amplitude decay in the same
direction as the propagation direction. 
Regarding the effective indices in medium 2, it turns out that $K_2$ = 0.0252,
which is significantly larger that the medium value of $\kappa_2$ = 0.0015, but
that $N_2$ $\approx$ $\eta_2$ = 1.045.  The next example will involve more
significant changes in $N_2$, which is proportionate to the real part of the propagation 
vector and determines the dispersion relation.

\begin{figure}
\includegraphics[width=4in]{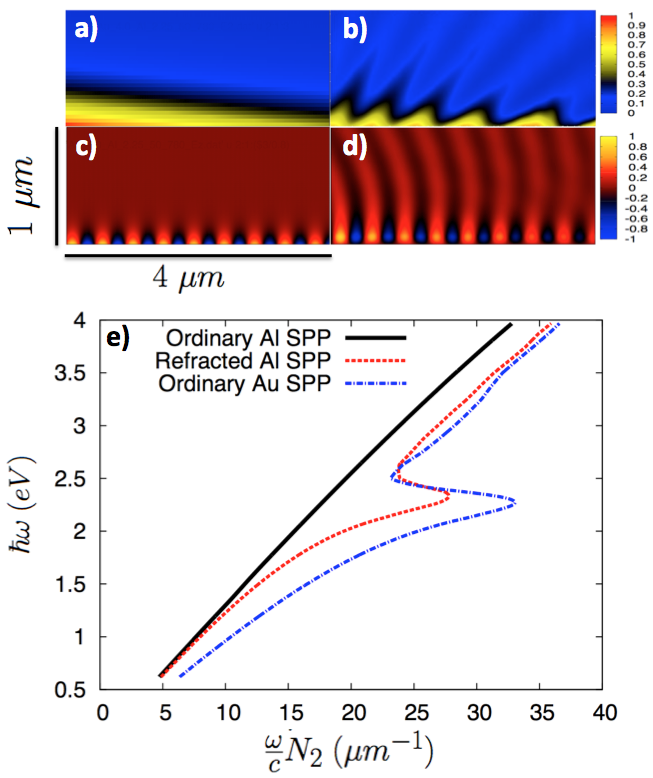}
\caption{
Field profiles of the refracted SPP on 
an Al surface arising from refraction of an incident SPP on a Au
surface. An $\epsilon^D_1$ = 4 material is assumed to be above the Au
surface and glass, $\epsilon^D_2$ = 2.25, is assumed to be above the
Al surface.  The incident SPP was excited with 780 nm light and 
had incident angle 50$^{\circ}$. 
(a) and (b): Analytical CSL and FDTD electric field intensities,
respectively.  (c) and (d): Analytical and FDTD instantaneous
electric field components, respectiely. 
(e) Dispersion relation for the SPP generated on Al (solid red) showing how
refraction imparts strong features of Au's SPP
dispersion (dashed blue)  onto the refracted wave.}
\label{fig:AuAl_Montage}
\end{figure}

As a second example consider $\lambda_0$ = 780 nm excitation of SPPs on Au as metal 1
($\epsilon_1 = -22.660 + 1.411 \: i$\cite{JC_PRB_1972}),
with a high refractive index material ($\epsilon^D_1$ = 4, e.g. TiO$_2$) above. 
These SPPs refract at an interface with aluminum as metal 2
($\epsilon_{2} = -66.263 + 45.719 \: i$\cite{Palik})
with glass ($\epsilon^D_2=2.25$) above. 
These result in medium refractive indices of
$\eta_1 = 2.202$, $\kappa_1 = 0.014$, $\eta_2 = 1.517$, and
$\kappa_2 = 0.012$.
In this example, when $\theta_1$ = 
50$^{\circ}$ $\theta_2$ $\approx$ 
90$^{\circ}$ while $\phi_2$ $\approx$ 1.5$^{\circ}$,
making this akin to total internal reflection (TIR) (see Fig. S4).  
FDTD results again show a high degree of similarity
in propagation direction and attenuation behavior (see Fig. 3 (a)-(d))
and a strong quantitative agreement in the predicted propagation
and confinement lengths (see Fig. S2). 
We focus only on the fields in medium 2, but again interference patterns
can be seen in the FDTD fields (Figs. 4 (c) and (e)) resulting from
interference of the reflected and incident SPP in medium 1.  Analysis of
the lines of constant phase and amplitude from FDTD simulations gives
a propagation direction of 90$^{\circ}$ and an attenuation
direction of approximately 0.5$^{\circ}$, respectively, in 
excellent agreement with CSL predictions.
Interestingly, some experimental evidence for this particular consequence
of Snell's law (TIR plasmons) 
has already been reported \cite{snell-2}.

The FDTD propagation length of the refracted
SPP is approximately
5.6 $\mu$m, and CSL also predicts 5.6 $\mu$m
(see Fig. S2).  The corresponding ordinary
SPP on an Al/glass surface would have a propagation length of
5.0  $\mu$m, so this provides an example of modest propagation
length enhancement.  
The refraction also produces an extremely confined mode:
FDTD $L_C$ 
$\approx 0.1 \mu$m; CSL $L_C$ =
0.08 $\mu$m (see Fig. S2).  
$L_P$ may be enhanced
proportionally to $N_2/\eta_2$, so it is expected the refracted
SPP will have a shorter wavelength than an ordinary SPP propagating
in medium 2.  Indeed, we find the refracted FDTD $\lambda_{SPP}$
to be 486 nm compared to a wavelength of
462 nm predicted by CSL. An ordinary SPP propagating
in medium 2 would
have a wavelength of 514 nm.
In this example both the real and imaginary parts of the effective
index of the refracted SPP are different from the medium 2 values:
$N_2$ = 1.688 compared to $\eta_2$ = 1.517 and $K_2$ = 0.738 compared to
$\kappa_2$ = 0.012.

It is also interesting for this example to consider the behavior of the refracted
SPP in  Al/glass across a spectrum of incident frequencies, i.e.
its dispersion.  Unlike an ordinary SPP dispersion in in an Al/glass system, the refracted
SPP dispersion also has information about  the refractive index, $\eta_1 + i \kappa_1$  and
the angle of incidence, $\theta_1$, in the Au/$\epsilon^D_2$ = 4 medium.
Mapping out the real part of the propagation vector, $(\omega / c) N_2$, 
with $\theta_1$ = 50$^{\circ}$ shows a startling departure from
the ordinary Al/glass result (Fig. 3 (e)).  The ordinary Al/glass
dispersion is relatively featureless, the wavevector increases uniformly with $\omega$
so that the SPP group velocity is approximately independent of frequency.
The dispersion of the refracted SPP shows markedly different features,
including a dramatic slowing of group velocity in the frequency range
between 1.5 and 2 eV
and a back-bending region between 2 and 3 eV.  These features mirror
dispersion characteristics of SPPs on the gold/$\epsilon^D_2=4$ medium (see Fig. 3(e)). 
To see the relationship between the details of the incident SPP and the dispersion
of the refracted SPP more clearly, we note that $\eta_2^4$ can be factored
out of b Eq.~\ref{eq:Root} so the root can be written
$N_2 = \frac{1}{2}\sqrt{a + \eta_2^2 \sqrt{1 + f}}$.  We can then
expand $\sqrt{1 + f}$ to first order in $f$, yielding the following approximation
of $N_2^2$,
\begin{equation}
N_2^2 \approx \eta_2^2 + \beta_1^2
- \frac{2}{\eta_2} \alpha_1 \beta_1 \kappa_2
+ \frac{1}{4 \eta_2^2}\left(\alpha_1^4 + \beta_1^4 + \kappa_2^4 + 2\left(\kappa_2^2 \alpha_1^2 - \kappa_2^2 \beta_1^2 + \beta_1^2 \alpha_1^2 \right)  \right).
\end{equation}
$\eta_2$ makes a strong contribution to $N_2$, which is to be expected
as the dispersion of the refracted SPP should depend on the material properties
of the metal/dielectric supporting its propagation.  What is interesting to note is the
fact the terms involving $\alpha_1$ ($\beta_1$) can make strong contributions
to $N_2$ when $\theta_1$ is large and $\eta_1$ ($\kappa_1$) is large.
Both $\eta_1$ and $\kappa_1$ tend to be large in the vicinity of the
surface plasmon resonance (SPR) of material 1, which in terms of the SPP
dispersion, is associated with the back-bending or anomalous dispersion
region~\cite{PRL_Ritchie_1973, AKB_PRL,PRB_Halevi, Lipson}.
Back-bending features can be encoded from material 1 to material
2 in general, but they will often be associated with high loss.  Recalling the
definitions of $L_P$ and $L_C$,  
we observe that the any additional loss
imparted by refraction leads to greater confinement of the SPP
and not additional attenuation in the propagation direction.

In this Letter, we presented a complex generalization of Snell's
law (CSL) to be applied to the refraction of SPPs propagating across a metal-metal
interface.  This theory predicts several surprising features of the refracted
SPPs which were validated using 3-D electrodynamics simulations.
In particular, we demonstrated that refraction can generate SPPs
that are inhomogeneous in the plane of propagation.  
It is possible to introduce significant confinement and, in certain cases, propagation
length enhancement to the refracted modes.  
A further consequence
is that the refracted SPPs obey unique dispersion relations that depend on the
supporting medium as well as the incident medium.  
The theory should also be applicable to more complicated structures such as
the layered wave guide modes 
described by Atwater and co-workers for the measurement
of negative refraction in the visible spectrum \cite{LDA_Science}, and indeed such 
structures might provide experimentally realizable 
systems for
the peculiar predictions that result from the CSL discussed here.
The theoretical and numerical results presented suggest that simple geometric principles 
such as the CSL discussed here can
offer novel
and powerful strategies for engineering SPPs, which could be particularly
useful for optoelectronics and sensing applications.
For example, although we do not discuss this application
in detail, FDTD simulations demonstrate SPP focusing can be achieved
using a metal region acting as a plasmonic lens (see Fig. S6).  The focusing
through refraction of such a lens could allow the generation of tightly
confined and controlled SPP modes without loss of propagation length.  
Optical switching devices could also be constructed using
incident angle or superstrate dielectric constant as nobs to induce
a dramatic change in the propagation behavior of the refracted SPP.  
Similarly, the dependence of the refracted wave on the dielectric properties
of the incident medium may also prove useful in chemical and biological sensing
applications. 

This work was performed, in part, at the Center for Nanoscale Materials, a U.S. Department of Energy, Office of Science, Office of Basic Energy Sciences User Facility under Contract No. DE-AC02-06CH11357.
JM and GCS were supported by the Department of Energy, Office of Basic Energy Science, 
under grant DE-FG02-10ER16153.

\bibliography{Snell_v21}
\end{document}


\title{Supplementary Material for: The Refraction of Surface Plasmon Polaritons}

\author{Jonathan J. Foley IV, Jeffrey M. McMahon, George C. Schatz, Stephen K. Gray }

\maketitle

\tableofcontents
\listoffigures



\section{Propagation and confinement lengths of the refracted SPPs}

Figures S1 and S2 show the complex Snell's law (CSL) and FDTD electric field intensities.
The CSL propagation length ($L_P$) and confinement
length ($L_C$) can be computed from the formulae given in the text,
and depend on the refraction angles $\theta_2$ and $\phi_2$ as well
as the imaginary part of the effective medium index, $K_2$. 
To extract
the same quantities from the FDTD results, we 
sample electric field 
intensities along the propagation direction and in the direction perpendicular
to the propagation direction.
Exponentials of the form $ | {\bf E}_0 | ^2 \; {\rm exp}(-r/L_i)$ are
fit to the sampled data to extract $L_P$ and $L_C$, where $r$ is 
the distance along the
appropriate scan direction.  Note Fig. S1 we sample approximately
5 $\mu$m along $\hat{{\bf a}}$ starting near the interface.
While longer sampling may seem desirable, care must be taken to avoid
sampling near the PMLs where the fields are rapidly attenuated and
simulations with significantly larger computational domains becomes
extremely challenging.  Calculations with ordinary SPPs
have verified that the attenuation behavior in the first few
$\mu$m of propagation is generally representative of the SPP
attenuation behavior.  The figures and captions show
that the FDTD results are reasonably well described by exponentials
and that the resulting propagation lengths are in good accord with the
CSL predictions.

\section{Dependence of refraction angles and effective indices on incident angle}

The angles of refraction ($\theta_2$ and $\phi_2$), as well as the effective SPP 
index ($N_2 + iK_2$), depend on the 
incident angle ($\theta_1$) via the CSL of Eqs. (5) and (6).  
A further subtle point is that sin$^{-1} \theta_1$ required for their
evaluation is multivalued, and 
$\theta_2$ and $\phi_2$ each have two possible values:  the principal value,
which will be between 0 and $\pi/2$, and what we will call the {\it secondary value},
which will be between $\pi/2$ and $\pi$.  The secondary value is related
to the principal value simply by $\theta^{SV} = \pi - \theta^{PV}$,
where the superscript PV (SV) denotes the principal (secondary) value.
Importantly, not all possible values of $\theta_2$ and $\phi_2$ satisfy
the condition $N_2 K_2 \: {\rm cos}(\theta_2 - \phi_2) = \eta_2 \kappa_2$ 
(Eq. (9) in the text).
To resolve the apparent ambiguity in assignment of the angles of refraction,
we note that for normal incidence ($\theta_1 = 0$),
both $\theta_2$ and $\phi_2$ must be zero, so $\theta_2^{PV}=0$
and $\phi_2^{PV}=0$ is the only possible angle pair.  For incident
angles larger than 0, we assign the angles by choosing
the angle pair that ensures that (a) Eq. (9) is satisfied
and (b) both $\theta_2$ and $\phi_2$ vary smoothly as a function
of $\theta_1$, starting from $\theta_1=\theta_2=\phi_2=0$.    

The trajectories of $\theta_2$ and $\phi_2$, along
with $N_2$ and $K_2$, as a function of
$\theta_1$ for the two refraction examples discussed in the text
are shown in Fig. S3 and S4.  (Note that if one expands the
scales in Fig. S4 in the $\theta_1$ = 40-50$^\circ$ range,  
the corresponding functions are actually smoothly varying.)

Terms from Eq. (12) of the main text contributing 
strongly to the dispersion encoding of gold
onto the aluminum surface are plotted in Fig. S5. (Terms
that are negligible across the frequency range are omitted
from the plot.)  We note that $\beta_1^2$ is similar
in magnitude to $\eta_2^2$ in the back-bending region.  This is important
because $\beta_1$ (through $\kappa_1$) will typically be 
more sensitive to the dielectric above metal 1 than $\eta_2$ will
be to the dielectric above metal 2.  Hence, the leading
$\beta_1$ term in the expansion of $N_2$ leads to an enhanced
sensitivity to the dielectric environment above metal 1, which
could be advantageous for sensing applications.

Although somewhat peripheral to the discussion, it should be noted that other
phenomena based on refraction, such as focusing by a lens, have analogs with absorbing material 
and can be treated with CSL.  
Fig. S6 demonstrates this with an SPP excited on a Al surface with a high-refractive
index superstrate (e.g. diamond, $\epsilon=6$) incident up
a silver cylindrical lens, also coated with a superstrate with $\epsilon=6$.
At this frequency, we arrive at medium refractive indices
$\eta_1 = 2.69, \kappa_1 = 0.075, \eta_2 = 4.08$, and $\kappa_2 = 0.064.$  
The ratio of $\eta_1$ to $\kappa_1$ is 0.66, 
analogous to geometric optics refraction at an air/glass interface. 
Hence, for small incident angles we expect similar focusing
behavior as a glass lens in air, which is qualitatively seen in 
Fig. S6.

\section{Finite-difference time-domain calculations}
The complex Snell's law predictions were validated by numerical solution
of Maxwell's equations using 3-D finite-difference time-domain
(FDTD) calculations~[1,2].  The structures described in the paper (see Fig. 1(a) of
the main text) are represented on a 
uniform rectangular grid with spacings dx = dy = dz = 5 nm. 
A line of point dipoles situated just above metal 1 in its dielectric
region is used to excite the incident SPPs.
The actual calculations, employing our own 3-D FDTD code, were most conveniently
accomplished by rotating the structure, as opposed to the source, in
order to achieve various angles of incidence, a particular example
being indicated  in Fig. S7.
The calculated results are then rotated to be consistent
with the coordinate system in Fig. 1 of the main text.
The overall size
of the computational domain for a typical calculation is 
(x,y,z) = 
(8$\mu m$, 14$\mu$m, 0.725$\mu$m).  We crop the field maps 
to highlight the behavior of the refracted SPPs over the normal propagation of the
incident SPP.  Fig. S7 represents a field map without rotation or cropping.
The metal slabs in all simulations have a thickness of 500 nm (in the z-direction) with a 40 nm dielectric
layer above and 35 nm layer below the metal; a 75 nm thick pml is used above and below
the dielectric layers.  The metal permittivities are modeled using a 
Drude + two Lorentz form~[2], with coefficients specifically fit for each example 
to closely reproduce the experimental permittivity values cited in the main text.
The SPP excitation source is modeled by a line of point dipoles,
positioned 10 nm above the surface of metal 1 (z = 620 nm),
and spanning the width of metal 1 (approximately 8$\mu$m),  
each dipole oscillating at the desired
excitation frequency.
Perfectly matched layers~[1] are used at the boundaries of the computational
domain. 
The FDTD simulation evolves the electric (and magnetic) fields with a time-step of approximately
1.8 x 10$^{-17}$ s for a total time of 100 fs; for the computational domains considered,
such a simulation typically takes about one hour on 120 processors.

To image the surface wave propagation, 
the $z$ component of the FDTD electric field, 
$E_z \left( t \right)$, is Fourier transformed on the excitation frequency $\omega$ to obtain  
the frequency domain result,
$E_z \left( \omega \right)$. 
(Other components yield similar results.) 
This represents the complex phasor field at this frequency that is associated with an
actual, real electric field time-dependence given by 
$E_z (t) = \mathrm{Re} \left[ E_z \left( \omega \right) \exp \left( -i \omega t \right) \right]$. 
An instantaneous field or a snapshot, which contains phase information, is obtained by evaluating
this latter expression for some arbitrary value of $t$ which we take to be $t$ = 0.
The associated field intensity is the time average of the square of the 
instantaneous field,
which is of course $| {\bf E}_z|^2$. 
For simplicity, the spatial dependence has been suppressed in the above discussion.
Generally, the fields are evaluated for in a plane of x,y values at a z value
corresponding to about 120nm above the metal surfaces.

\begin{enumerate}

\item A. Taflove and S. C. Hagness, {\it Computational Electrodynamics: the finite-difference 
time-domain method} 2$^{{\rm nd}}$ ed. (Artech, 2000). 

\item J. M. Montgomery, T.-W. Lee, and S. K. Gray, J. Phys.: Conds. Mater {\bf 20},
323201 (2008).
 
\end{enumerate}

\newpage
\begin{figure}[!ht]
\includegraphics[width=5.5in]{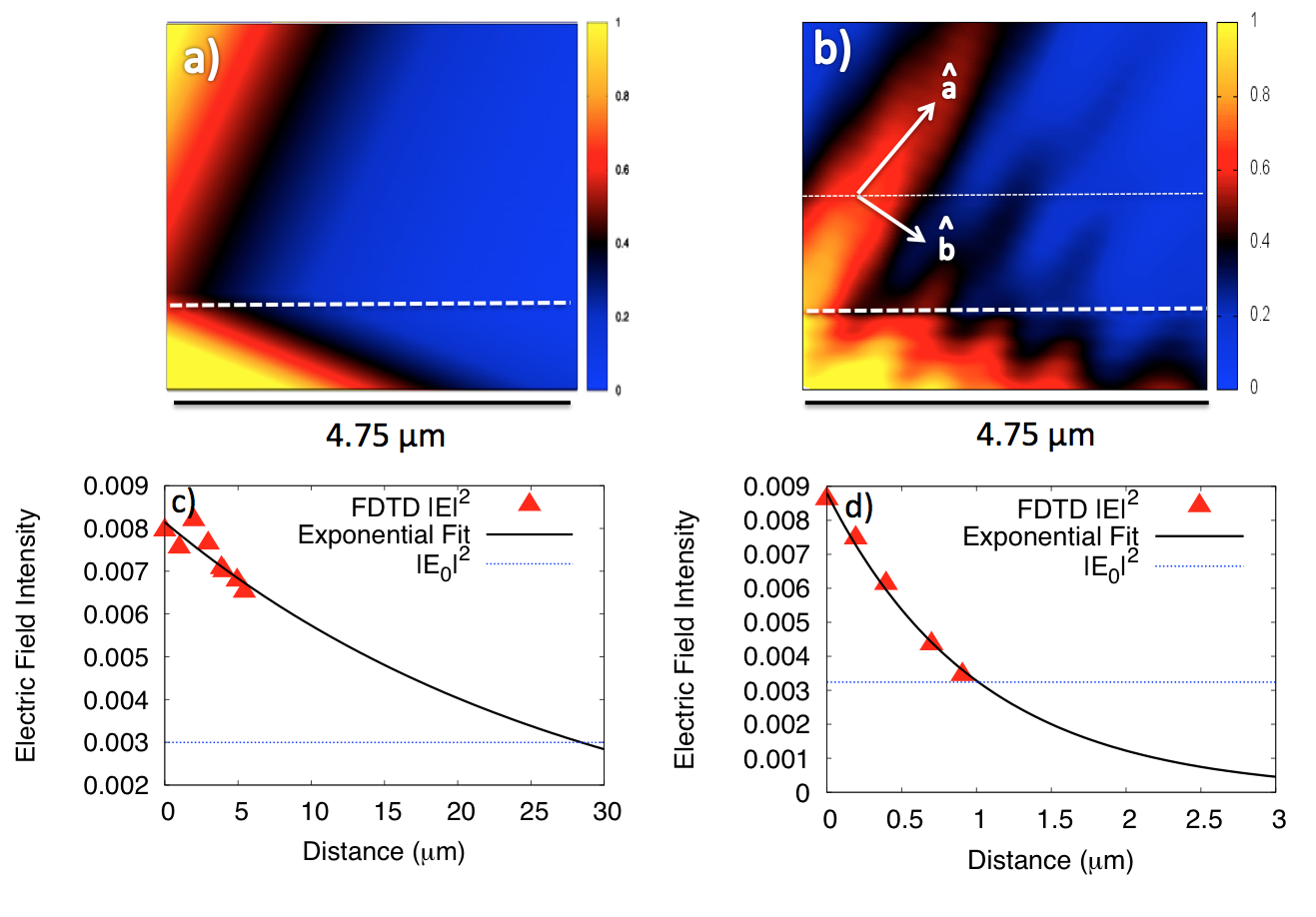}
\caption[Propagation and Confinement Length (Au/Ag) ]{ (a) and (b) : 
Electric field intensities, $|{\bf E}|^2$, for an  SPP on
Au (below dashed white line), being refracted onto Ag 
(above white dashed line). The angle of incidence is 25$^{\circ}$,
the vacuum wavelength is 532 nm, and air is assumed to be above both
metals. 
The CSL result is in panel (a) and the rigorous electrodynamics
calculation, via the FDTD method, is in panel (b).
(c) The propagation length is calculated by
evaluating $ | {\bf E} | ^2$ from the electrodynamics calculation
along the vector $\hat{{\bf a}}$
and fitting an exponential function.
CSL predicts that the vector $\hat{{\bf a}}$ will point approximately
27$^{\circ}$ from the normal to the interface at the white dashed line.
The exponential, solid curve, crosses $ | {\bf E}_0 | ^2/e$
at 28 $\mu$m, in excellent agreement analytical prediction of
27 $\mu$m from CSL. (d)
The confinement length is calculated by
evaluating $ | {\bf E} | ^2$ from the electrodynamics calculation
along the vector $\hat{{\bf b}}$
and fitting an exponential function to the field values.  CSL
predicts that the vector $\hat{{\bf a}}$ will point approximately 114$^{\circ}$
to the normal to the interface.
The exponential, solid curve, crosses $ | {\bf E}_0 | ^2/e$
at 1.1 $\mu$m, also in good agreement analytical prediction of
1.7 $\mu$m from CSL.}
\label{fig:AuAg_Montage_E2}
\end{figure}

\begin{figure}[!ht]
\includegraphics[width=5.5in]{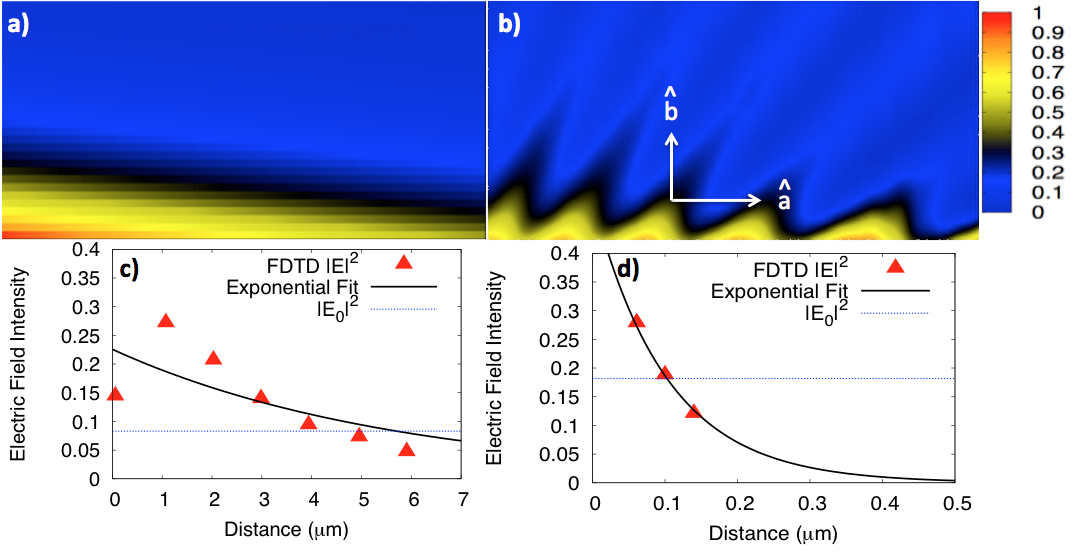}
\caption[Propagation and Confinement Length (Au/Al)]{
(a) and (b): 
Electric field intensities, $|{\bf E}|^2$,  resulting from an  SPP on
Au being refracted onto Al. 
The angle of incidence is 50$^{\circ}$,
the vacuum wavelength is 780 nm.
An $\epsilon^D_1 = 4$ dielectric is assumed
to be above the Au surface, and glass ($\epsilon^D_2 = 2.25$) 
is assumed to be above the Al surface. 
Only the refracted (on Al) portion is displayed.
The CSL result is in  (a) and from rigorous electrodynamics
(FDTD) calculation is in (b).
(c) The propagation length is calculated by
evaluating field intensity from the electrodynamics calculation
along the vector $\hat{{\bf a}}$
and fitting an exponential function.
CSL predicts that the vector $\hat{{\bf a}}$ will point approximately
90$^{\circ}$ from the normal to the interface.
The exponential, plotted in (c), crosses $ | {\bf E}_0 | ^2/e$
at 5.6 $\mu$m, in excellent agreement analytical prediction of
5.6 $\mu$m from CSL.
The confinement length is calculated by
evaluating $ |  {\bf E} | ^2$ from the electrodynamics calculation
along the vector $\hat{{\bf b}}$
and fitting an exponential function to the field values.  CSL
predicts that the vector $\hat{{\bf a}}$ will point approximately 1$^{\circ}$
to the normal to the interface.
The exponential, plotted in (d), crosses $ |  {\bf E}_0 | ^2/e$
at 0.1 $\mu$m, also in agreement analytical prediction of
0.08 $\mu$m from CSL.}
\label{fig:AuAl_Montage}
\end{figure}

\begin{figure}[!ht]
\includegraphics[angle=270,width=3.in]{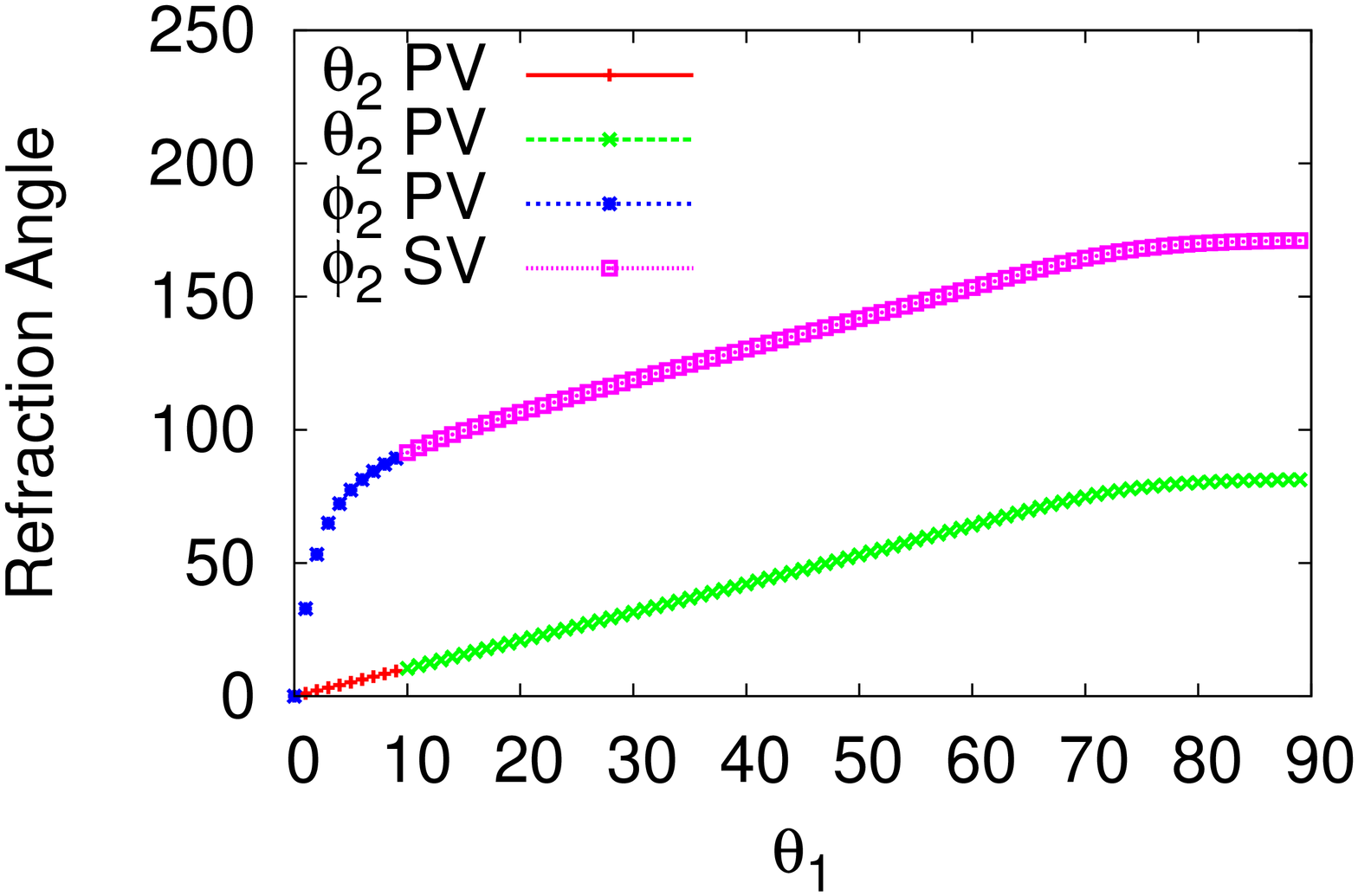}
\includegraphics[angle=270,width=3.in]{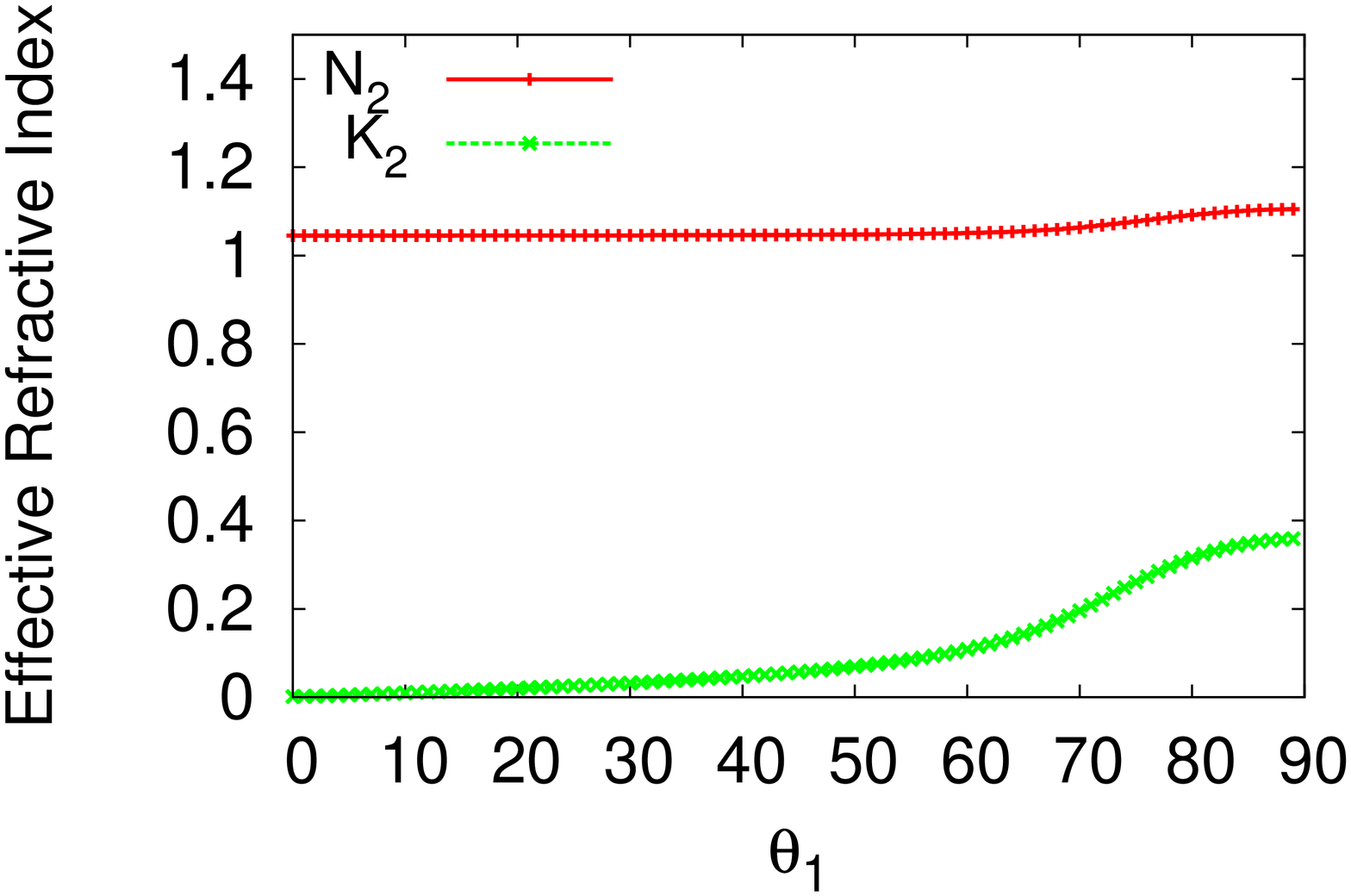}
\caption[Trajectory of Refraction Angles and Indices (Au/Ag)]{(Left) Trajectory of refraction angles $\theta_2$ and $\phi_2$ as a function
of $\theta_1$ for an SPP excited by 532 nm light on
a gold surface with air above (medium 1) and refracting onto an aluminum surface with glass above (medium 2).
For incident angles larger than 9$^{\circ}$, the principal values of $\theta_2$ and $\phi_2$
do not satisfy Eq. (9).  The angle pair ($\theta_2^{PV},\phi_2^{SV}$) maintains
continuity beyond $\theta_1 = 9^{\circ}$.  (Right) Trajectory
of effective refractive indices for the refracted SPP as a function
of $\theta_1$. }
\label{fig:AuAg_Ang}
\end{figure}

\begin{figure}[!ht]
\includegraphics[angle=270,width=3.in]{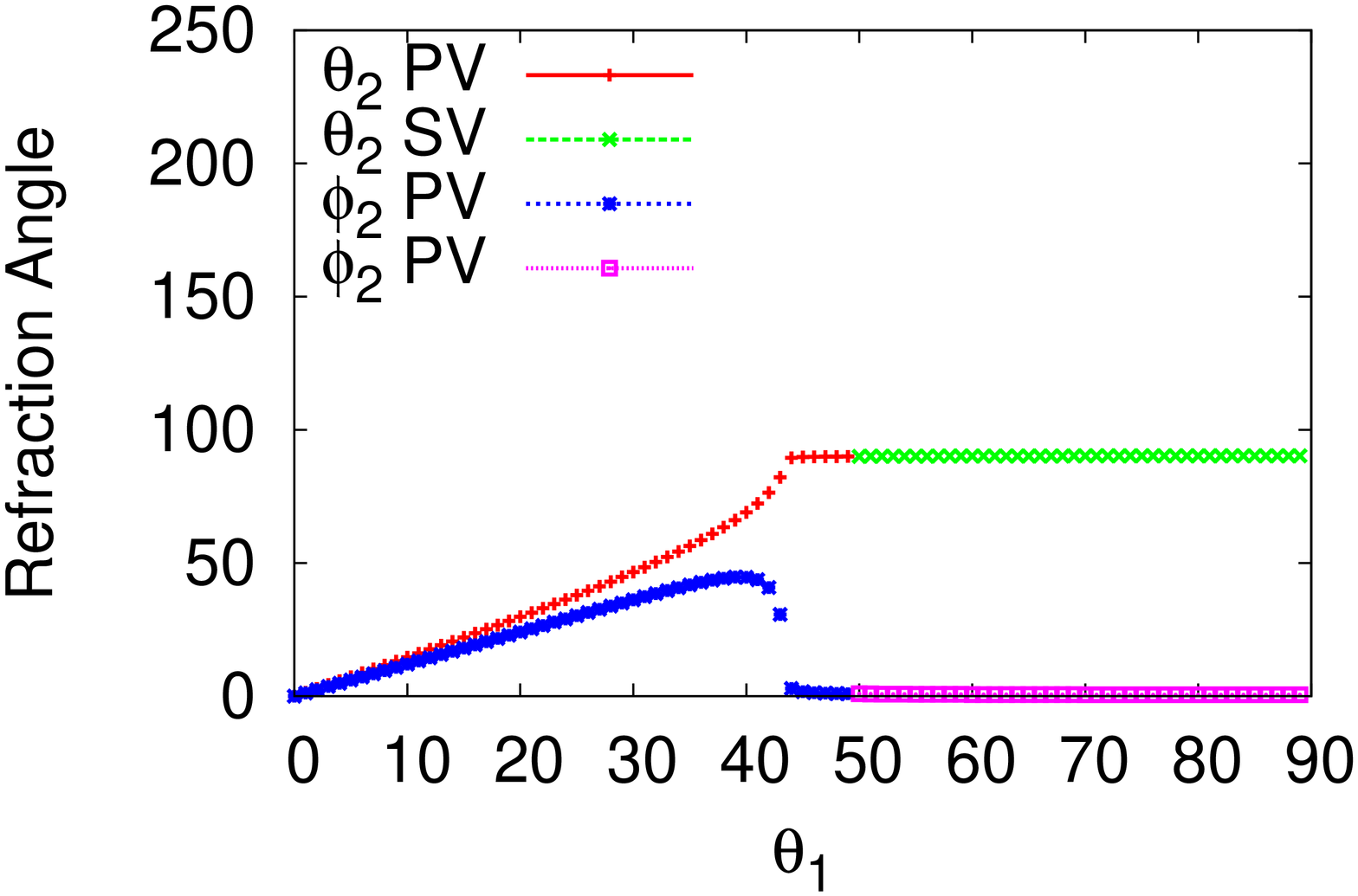}
\includegraphics[angle=270,width=3.in]{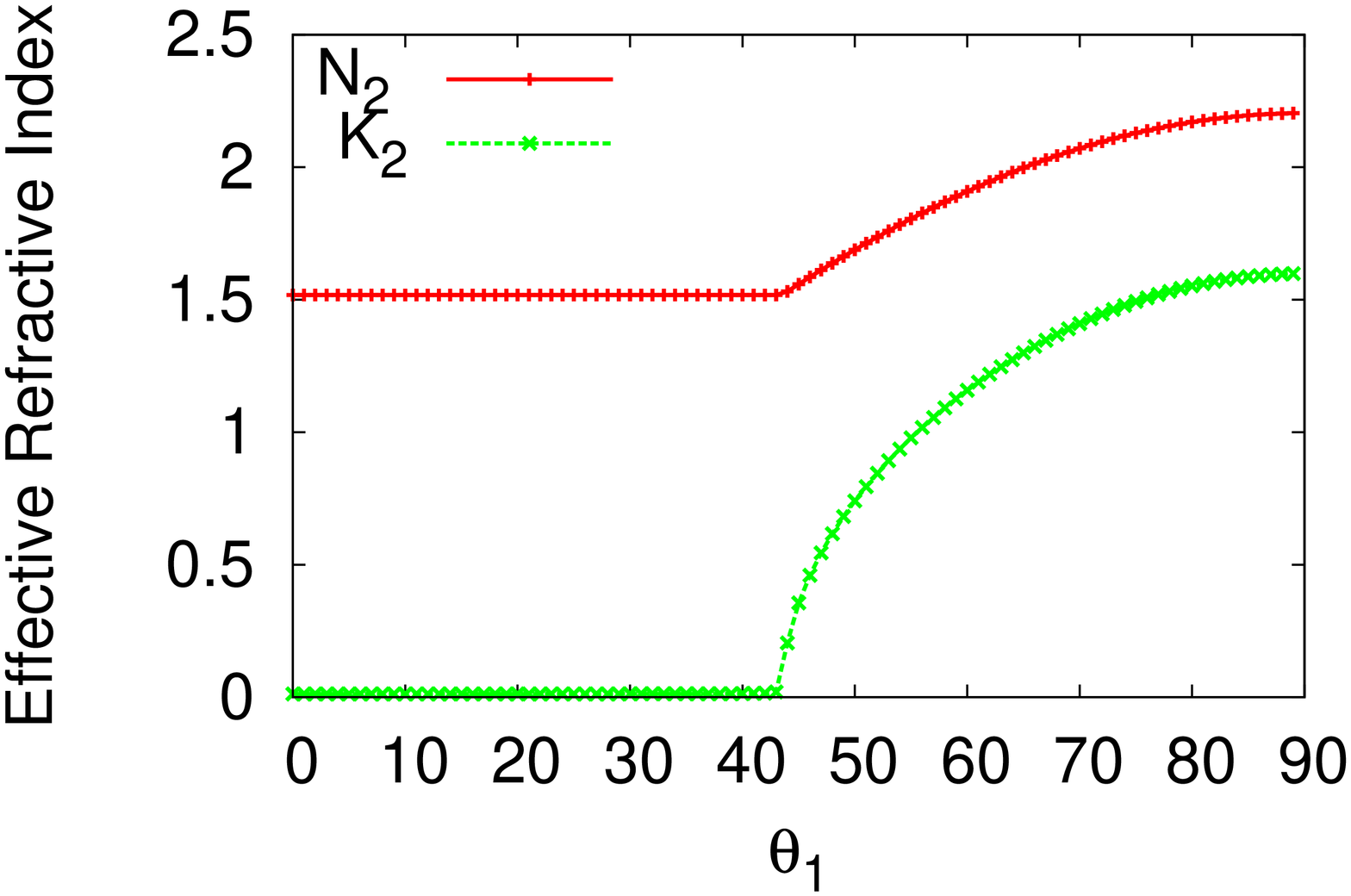}
\caption[Trajectory of Refraction Angles and Indices (Au/Al)]{(Left) Trajectory of refraction angles $\theta_2$ and $\phi_2$ as a function
of $\theta_1$ for an SPP excited by 780 nm light on
a gold surface with $\epsilon=4$ above (medium 1) and refracting onto an aluminum surface with glass above (medium 2).
For incident angles larger than 50$^{\circ}$, the principal values of $\theta_2$ and $\phi_2$
do not satisfy Eq. (9).  The angle pair ($\theta_2^{SV},\phi_2^{PV}$) maintains
continuity beyond $\theta_1 = 50^{\circ}$.  (Right) Trajectory
of effective refractive indices for the refracted SPP as a function
of $\theta_1$. Dramatic departure of both effective indices from the material
indices can be seen starting at $\theta_1 = 45^{\circ}$.  This strong departure
can also be seen in the imprinted dispersion in Fig. 3 in the main text.}
\label{fig:AuAl_Ang}
\end{figure}

\begin{figure}[!ht]
\includegraphics[angle=270,width=5.in]{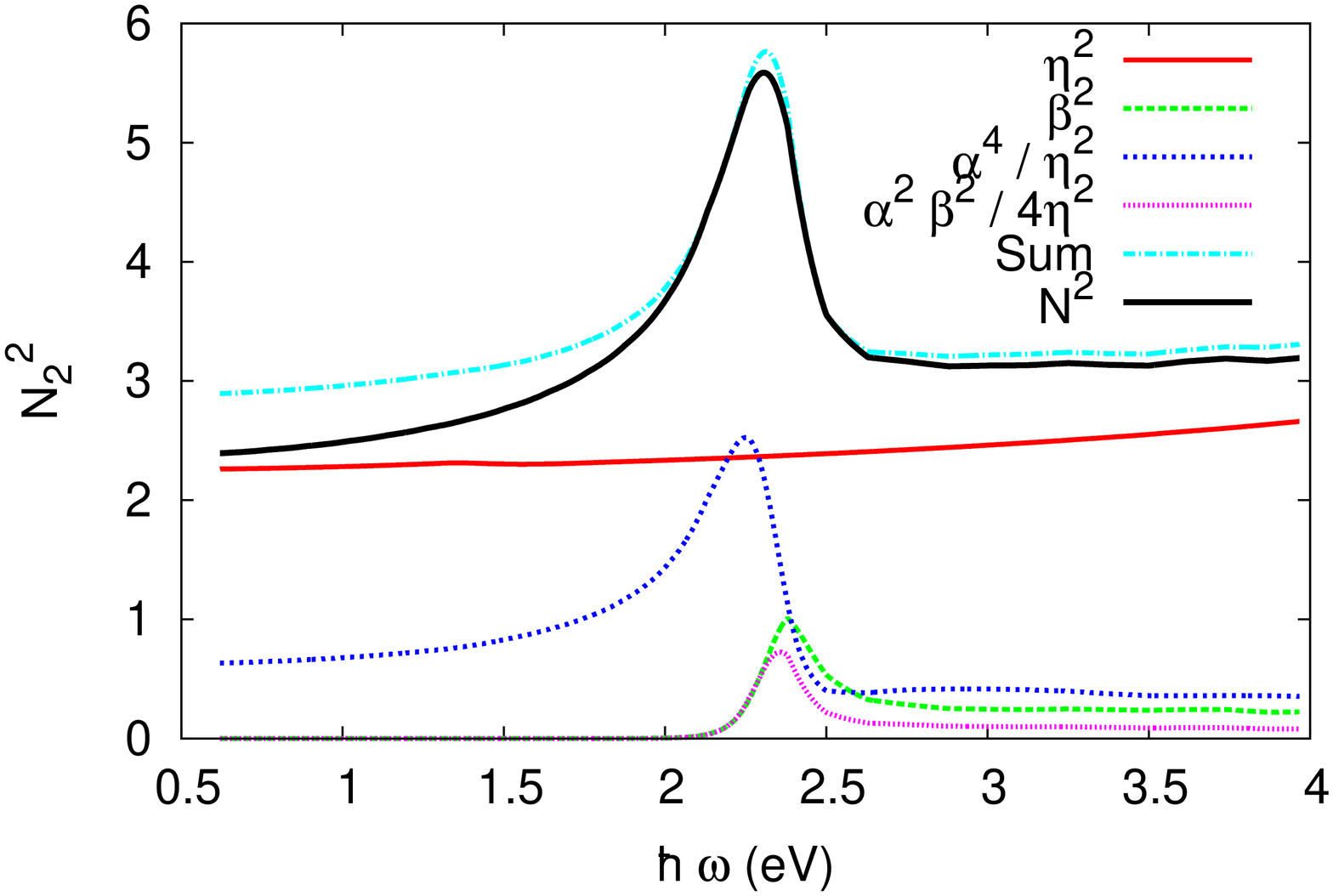}
\caption[Imprinted dispersion terms]{An approximate expansion of $N_2$ is given in the text.
Here we see the contribution of the largest terms to $N_2$, including
$\eta_2$, $\alpha_1$, and $\beta_1$.  (For simplicity of presentation, the
figure legend uses $N$ = $N_2$, $\alpha$ = $\alpha_1$, $\beta$ = $\beta_1$
and $\eta$ = $\eta_2$.)
Terms from the expansion in Eq. 12
in the text are omitted from the plot if they are negligible relative to $\eta_2$
across all frequencies considered.}
\label{fig:AuAl_Expansion}
\end{figure}

\begin{figure}[!ht]
\includegraphics[width=5in]{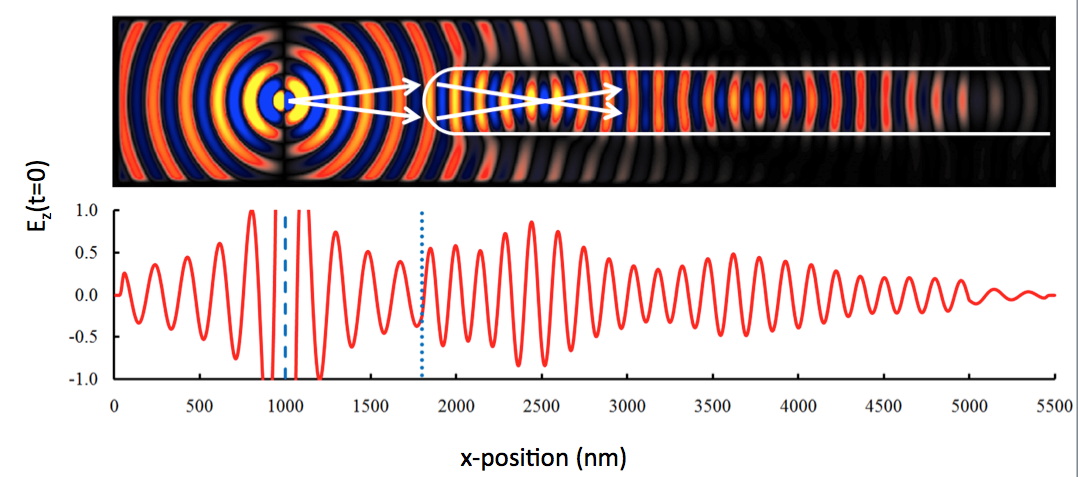}
\caption[Focusing by a plasmonic lens]{FDTD calculations of an SPP 
excited by 495 nm light on an Al surface,
and focused by a cylindrical Ag lens region. A $\epsilon$ = 6 dielectric
material is assumed to be above both regions. Upper image: snapshot of
Re($E_z$) 10 nm
above the film surface. The Ag film is outlined in white. Lower image: 
slice through the middle of the upper image. The blue dashed
line indicated the dipole position, and the blue dotted line indicates the
surface of the Ag film.}
\end{figure}

\begin{figure}[!ht]
\includegraphics[width=5in]{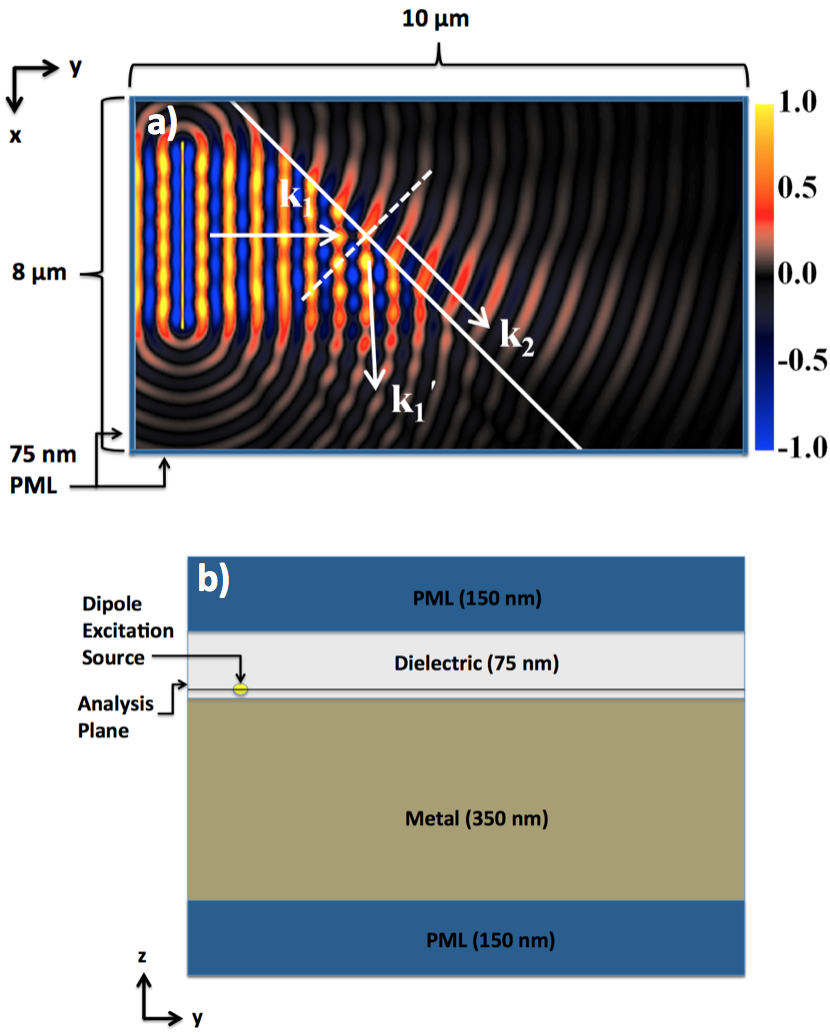}
\caption[Illustrative figure showing FDTD calculation of SPP refraction]{Illustrative figure showing FDTD
calculation of SPP refraction.  
(a) is a 2-D cut of the instantaneous field,  ${\rm Re}(E_z)$,
in the x-y plane at a z value corresponding to 10 nm above the metal surfaces. 
The excitation source
(an $\approx$ 5 $\mu$m line of point dipoles also 10 nm above the metal surface)  
is evident on the left, and generates an SPP with 
wavevector ${\bf k_1}$ incident upon an angled interface between medium 1 and medium 2
(solid white line).
The computational domain
is typical of the calculations presented in the main text - spanning 8$\mu$m in
x, 10$\mu$m in y, and 0.725$\mu$m in z.
The SPP is reflected (${\bf k_1'}$) and refracted (${\bf k_2}$) at the interface.
The actual example shown corresponds to medium 1 being Ag, medium 2 being Al, $\epsilon$ = 6
dielectric above each metal, 
and 496 nm incident light, which yields a TIR-like SPP wave in medium 2.  The the outter 75 nm of
the domain consists of perfectly matched layers (PML).
(b) is a schematic representation of an y-z slice of the computational domain, 
showing the 350 nm metal slab with 75 nm dielectric layer above as well as 150 nm PMLs on top and bottom.
The approximate z-positon of the line of dipoles and the analysis plane (where the fields are imaged) is 
also indicated.
}

\end{figure}